\documentstyle[preprint,tighten,aps]{revtex}
\begin{document}

\def\beqar{\begin{eqnarray}}
\def\eeqar{\end{eqnarray}}
\def\be{\begin{eqnarray}}
\def\ee{\end{eqnarray}}
\def\beqast{\begin{eqnarray*}}
\def\eeqast{\end{eqnarray*}}
\def\be{\begin{enumerate}}
\def\ee{\end{enumerate}}
\def\lag{\langle}
\def\rag{\rangle}
\def\fnote#1#2{\begingroup\def\thefootnote{#1}\footnote{#2}
\addtocounter{footnote}{-1}\endgroup}
\def\beq{\begin{equation}}
\def\eeq{\end{equation}}
\def\haf{\frac{1}{2}}
\def\pa{\partial}
\def\ca{{\cal A}}
\def\cb{{\cal B}}
\def\plb#1#2#3#4{#1, Phys. Lett. {\bf B#2}, #3 (#4)}
\def\npb#1#2#3#4{#1, Nucl. Phys. {\bf B#2}, #3 (#4)}
\def\prd#1#2#3#4{#1, Phys. Rev. {\bf D#2}, #3 (#4)}
\def\prl#1#2#3#4{#1, Phys. Rev. Lett. {\bf #2}, #3 (#4)}
\def\mpl#1#2#3#4{#1, Mod. Phys. Lett. {\bf A#2}, #3 (#4)}
\def\rep#1#2#3#4{#1, Phys. Rep. {\bf #2}, #3 (#4)}
\def\llp#1#2{\lambda_{#1}\lambda'_{#2}}
\def\lplp#1#2{\lambda'_{#1}\lambda'_{#2}}
\def\slash#1{#1\!\!\!\!\!/}
\def\rp{\slash{R_p}~}

%%%%%%%%%%%%%%%%%%%%%%%%%%%%%%%%%%%%%%%%%%%%%%%%%%%%%%%%%%%%%%%%%%%%%%%%%%

\draft
\preprint{
\begin{tabular}{r}
KAIST-TH 97/04
\\
hep-ph/9704213
\end{tabular}
}
\title{
$R$-parity Violation and Semileptonic Decays of $B$-meson
}
\author{
Ji-Ho Jang 
\thanks{E-mail: jhjang@chep6.kaist.ac.kr},
Yeong Gyun Kim
\thanks{E-mail: ygkim@chep6.kaist.ac.kr},
and
Jae Sik Lee
\thanks{E-mail: jslee@chep6.kaist.ac.kr}
}
\address{
Department of Physics, Korea Advanced Institute of Science and
Technology \\
Taejon 305-701, Korea \\
}
%and \\
%$^{\rm{b}}$Department of Physics and Astronomy,
%The Johns Hopkins University \\
%Baltimore, Maryland 21218, USA.
%}

\maketitle

\begin{abstract}
We investigate the effects of $R$-parity violation on the semileptonic decays of
$B$-meson in the minimal supersymmetric
standard model with explicit $R$-parity violation and discuss its
physical implications. 
We find that the semileptonic decays of $B$-meson
can be largely affected by $R$-parity violation.
\end{abstract}


\newpage
%\section{introduction}
One of the most important objects of future experiments is to find
supersymmetry or a hint for it.
In supersymmetric models, there are gauge invariant interactions which
violate the baryon number $B$ and the lepton number $L$ generally. To prevent
presence of these $B$ and $L$ violating interactions in supersymmetric
models, an additional global symmetry is required. This requirement leads to
the consideration of the so called $R$-parity.
The $R$-parity is given by the relation $R_p=(-1)^{(3B+L+2S)}$ where 
$S$ is  the intrinsic spin of a field.
%In the supersymmetric standard models, the interactions which violate 
%the $R_p$ are generally not forbidden by the gauge invariance.
Even though the requirement of $R_p$ conservation gives a theory
consistent with
present experimental searches, there is no good theoretical justification
for this requirement. Therefore
models with explicit $R_p$ violation ($\slash{R_p}~$) 
have been considered by many authors \cite{ago}. 

In models without $R_p$, the supersymmetric
particles can decay into the ordinary particles alone. 
So the couplings which violate $R_p$ can be detected by
using the usual particle detectors.
If we discover a sign of $\slash{R_p}~$ in future experiments, it
may provide us with some hints for supersymmetry.
Among future experiments,
the upcoming experiments on $B$-mesons (BaBar, BELLE, HERA B, CLEO, RUN II at
FNAL) \cite{Bfac} motivate the study of the effects of $\slash{R_p}~$ on the
decays of $B$-meson.

In this paper, we study the semileptonic decays of
$B$-meson in the minimal supersymmetric
standard model (MSSM) with $\rp$. 
We investigate how much $\rp$ affects the semileptonic decay rates within
the present bounds and discuss its physical implications.
We find that the semileptonic decays of $B$-meson can be 
largely affected by $\slash{R_p}~$.

In the MSSM
the most general $R_p$ violating superpotential is given by
\beq
W_{R\!\!\!\!/_p}=\lambda_{ijk}L_iL_jE_k^c+\lambda_{ijk}'L_iQ_jD_k^c+
\lambda_{ijk}''U_i^cD_j^cD_k^c.
\eeq
Here $i,j,k$ are generation indices and we assume that possible bilinear terms
$\mu_i L_i H_2$ can be rotated away.
$L_i$ and $Q_i$ are the $SU(2)$-doublet lepton and the quark superfields and
$E_i^c,U_i^c,D_i^c$ are the singlet superfields respectively. 
$\lambda_{ijk}$ and
$\lambda_{ijk}''$ are antisymmetric under the interchange of the first two and
the last two generation indices respectively; $\lambda_{ijk}=-\lambda_{jik}$ and
$\lambda_{ijk}''=-\lambda_{ikj}''$. So the number of couplings is 45 (9 of the
$\lambda$ type, 27 of the $\lambda'$ type and 9 of the $\lambda''$ type).
Among these 45 couplings, 36 couplings are related with the lepton
flavor violation.
Usually, the constraints on the couplings with heavy fields are 
not as strong as those with light fields.

%\section{present bounds}
There are upper bounds on a {\it single} $R_p$ violating coupling from
several different sources \cite{han,beta,numass,agagra,Zdecay}.
Among these, upper bounds from atomic parity violation and $eD$ asymmetry
\cite{han}, $\nu_{\mu}$ deep-inelastic scattering \cite{han},
neutrinoless double beta decay \cite{beta}, $\nu$ mass \cite{numass}, 
$K^+,t-$quark decays \cite{agagra,choroy}, and $Z$ decay width \cite{Zdecay}
 are strong.
Neutrinoless double beta decay gives
$\lambda'_{111}<4\times10^{-4}$.
The bounds from $\nu$ mass are $\lambda_{133}<10^{-3}$
and $\lambda'_{133}<10^{-3}$.
From $K^+$-meson decays one obtain $\lambda'_{ijk}<0.012$ for $j=1$ and 2.
Here all masses of scalar partners which mediate the processes are assumed
to be 100 GeV.
Extensive reviews of the updated limits on a single $R_p$ violating
coupling can be found in \cite{bha,chahu}.

There are more stringent bounds on some products of
the $R_p$ violating couplings from the mixings of the
neutral $K$- and $B$- mesons and rare leptonic decays of
the $K_L$-meson, the muon and the tau \cite{choroy}, $B^0$ decays
into two charged leptons \cite{Lee}, $b\bar{b}$ productions at LEP \cite{Feng}
and muon(ium) conversion, and $\tau$ and $\pi^0$ decays \cite{Ko}.
%Some parts of upper bounds on $\rp$ couplings from both semileptonic and
%purely leptonic B meson decays were first given in Ref. \cite{Feng}.

$\rp$ couplings with heavy generation indices are only moderately constrained.
This means that we could constrain $\rp$ couplings with heavy generation 
indices from $B$ meson decays and $\rp$ signals could be found in the upcoming
experiments on $B$ mesons \cite{Bfac}. First examples on $\rp$
couplings from $B$ decays were given in Ref. \cite{Feng}.

In this paper we assume that the $B$ violating couplings $\lambda''$ 
are vanishing to avoid too fast proton decays.  Especially in the models with
a very light gravitino ($G$) or axino ($\tilde{a}$), $\lambda''$ have to be
very small independently of $\lambda'$ 
from the proton decay $p\rightarrow K^+G~({\rm or}~K^+\tilde{a})$ 
; $\lambda''_{112}<10^{-15}$ \cite{Choi}. 
One can construct a grand unified model which has only lepton number
non-conserving trilinear operators in the low energy superpotential when
$R_p$ is broken only by bilinear terms of the form $L_i H_2$
\cite{hasu}. And usually it may be very difficult to discern signals of
$B$-violating interactions above QCD backgrounds \cite{numass}.

%\section{four-fermion interactions}
In the MSSM with $R_p$, the terms in
the effective lagrangian relevant for the semileptonic $B$-meson decays are
\cite{GrLi}
\beq
{\cal L}^{eff}(b\rightarrow q e_l \bar{\nu}_l)=
-V_{qb}\frac{4G_F}{\sqrt{2}}\left[
(\bar{q}\gamma^{\mu}P_Lb)(\bar{e}_l\gamma_{\mu}P_L\nu_l)-
R_l(\bar{q}P_Rb)(\bar{e}_lP_L\nu_l)\right],
\eeq
where
$P_{R,L}=\frac{1}{2}(1\pm\gamma_5)$, 
%\beq
$R_l=r^2 m_{e_l} m_b^Y$ and
$r=\frac{\tan\beta}{m_{H^{\pm}}}$.
%\eeq
An upper index $Y$ denotes the running quark mass,
$\tan\beta$ is the ratio of the vacuum expectation values of the neutral
Higgs fields
and $m_{H^{\pm}}$ is the mass of the charged Higgs fields.
The first term in Eq. (2) gives the standard model (SM)
contribution and the second one
gives that of the charged Higgs scalars. Neglecting the masses of
the electron ($l=1$) and the muon ($l=2$), the contribution
of the charged Higgs scalars is zero. The contribution of the charged Higgs
scalars is not vanishing only when $l=3$ ;
$b\rightarrow q~\tau~\bar{\nu}_{\tau}$.
We neglect a term
proportional to $m_c^Y$ for $q=c$
since the term is suppressed by the mass ratio
$m_c^Y/m_b^Y$ and does not have the possibly large $\tan^2\beta$ factor.

In the MSSM without $R_p$, the exchange of the sleptons and the
squarks leads to the additional four-fermion
interactions which are
relevant for the semileptonic decays of $B$-meson.
Considering the fact that the CKM matrix $V$ is not an identity matrix,
the $\lambda'$ terms of the Eq. (1) are reexpressed in terms of the 
the fermion mass eigenstates as follow
\beq
W_{\lambda'}=\lambda'_{ijk}\left(N_iD_j-
\sum_{p}V^{\dagger}_{jp}E_iU_p\right)D^c_k,
\eeq
where $N_i$, $E_i$, $U_i$ and $D_i$ are the superfields with neutrinos,
charged leptons, up- and down-type-quarks and
$\lambda'$ have been redefined to absorb some field rotation effects. 
From Eq. (1) and Eq. (3) we obtain the effective interactions which are
relevant for the semileptonic decays of $B$-meson as follows
\beq
{\cal L}^{eff}_{\slash{R_p}}(b\rightarrow q e_l \bar{\nu}_n)=
-V_{qb}\frac{4G_F}{\sqrt{2}} \left[
{\cal A}_{ln}^q(\bar{q}\gamma^{\mu}P_Lb)(\bar{e}_l\gamma_{\mu}P_L\nu_n)-
{\cal B}_{ln}^q(\bar{q}P_Rb)(\bar{e}_lP_L\nu_n)\right],
\eeq
where we assume 
the matrices of the soft mass terms are diagonal in the fermion mass basis. 
Note that the operators in Eq. (4) take the same form as
those of the MSSM with $R_p$.  Comparing with the SM, the
above effective lagrangian includes the interactions even when $l$ and $n$ are
different from each other.
The dimensionless coupling constants
${\cal A}$ and ${\cal B}$ depend on the species of quark, charged lepton 
and neutrino and are given by
\beqar
{\cal A}_{ln}^q&=&\frac{\sqrt{2}}{4G_FV_{qb}}\sum_{i,j=1}^{3}
\frac{1}{2m^2_{\tilde{d}_i^c}}V_{qj}\lambda'_{n3i}\lambda'^*_{lji},
\nonumber \\
{\cal B}_{ln}^q&=&\frac{\sqrt{2}}{4G_FV_{qb}}\sum_{i,j=1}^{3}
\frac{2}{m^2_{\tilde{l}_i}}V_{qj}\lambda_{inl}\lambda'^*_{ij3},
\eeqar
where $q=u,c$ and $l$ and $n$ are the generation indices running from 1 to 3.

From the numerical values of \cite{PDG}, we find
\beqar
\ca_{ln}^u&=&\sum_{i=1}^{3}\lambda'_{n3i}
\left\{422\lambda'^*_{l1i}
~\left(\frac{V_{ud}/0.9751}{V_{ub}/0.0035}\right)
+96\lambda'^*_{l2i}
~\left(\frac{V_{us}/0.2215}{V_{ub}/0.0035}\right)
+1.52\lambda'^*_{l3i}\right\}
\left(\frac{100~ {\rm GeV}}{m_{\tilde{d}_i^c}}\right)^2,
\nonumber \\
\cb_{ln}^u&=&\sum_{i=1}^{3}\lambda_{inl}
\left\{1689\lambda'^*_{i13}
~\left(\frac{V_{ud}/0.9751}{V_{ub}/0.0035}\right)
+384\lambda'^*_{i23}
~\left(\frac{V_{us}/0.2215}{V_{ub}/0.0035}\right)
+6.1\lambda'^*_{i33}\right\}
\left(\frac{100~ {\rm GeV}}{m_{\tilde{l}_i}}\right)^2,
\nonumber \\
\ca_{ln}^c&=&\sum_{i=1}^{3}\lambda'_{n3i}
\left\{8.2\lambda'^*_{l1i}
~\left(\frac{V_{cd}/0.221}{V_{cb}/0.041}\right)
+36\lambda'^*_{l2i}
~\left(\frac{V_{cs}/0.9743}{V_{cb}/0.041}\right)
+1.52\lambda'^*_{l3i}\right\}
\left(\frac{100~ {\rm GeV}}{m_{\tilde{d}_i^c}}\right)^2,
\nonumber \\
\cb_{ln}^c&=&\sum_{i=1}^{3}\lambda_{inl}
\left\{32.7\lambda'^*_{i13}
~\left(\frac{V_{cd}/0.221}{V_{cb}/0.041}\right)
+144\lambda'^*_{i23}
~\left(\frac{V_{cs}/0.9743}{V_{cb}/0.041}\right)
+6.1\lambda'^*_{i33}\right\}
\left(\frac{100~ {\rm GeV}}{m_{\tilde{l}_i}}\right)^2.
\eeqar
Note the large numerical factors coming from the big differences between the
values of the CKM matrix elements.

%\section{decay rates}
When the species of the charged lepton and the neutrino are same,
the decay rate of the process $b\rightarrow q e_l \nu_l$ is
\beqar
\Gamma_{ll}^q&=&\frac{\left|V_{qb}\right|^2G_F^2m_b^5}{192\pi^3} 
%\nonumber \\
\left\{ \left| 1+\ca_{ll}^q\right|^2 \Gamma_W + \frac{1}{4}
\left|R_l+\cb_{ll}^q\right|^2\Gamma_H\right. \nonumber \\
&&\left.-2Re\left[(R_l+\cb_{ll}^q)(1+\ca_{ll}^{q*})\right]\frac{m_{e_l}}{m_b}
\Gamma_I\right\}.
\eeqar
And for the different species of the charged lepton and the neutrino,
the decay rate of the process $b\rightarrow q e_l \nu_n$ is
\beq
\left.\Gamma_{ln}^q\right|_{l\neq n}=
\frac{\left|V_{qb}\right|^2G_F^2m_b^5}{192\pi^3} 
\left\{ \left|\ca_{ln}^q\right|^2 \Gamma_W + \frac{1}{4}
\left|\cb_{ln}^q\right|^2\Gamma_H
-2Re\left(\cb_{ln}^q\ca_{ln}^{q*}\right)\frac{m_{e_l}}{m_b}
\Gamma_I\right\}.
\eeq
The subindices $W, H$ and $I$ of $\Gamma$ denote the $W$ mediated (SM),
charged Higgs mediated and interference contributions respectively. The
explicit forms and relations 
between $\Gamma_W$, $\Gamma_H$ and $\Gamma_I$ are given in
Ref. \cite{GrLi,Falk}. 
For $l=1$ and $2$, $R_l$ and the interference
term are vanishing assuming the electron and the muon
are massless.

%Neglecting the contribution from $b\rightarrow u$
%transition and the masses of the electron and the muon, 
%the $\Gamma_W$, $\Gamma_H$ and $\Gamma_I$ are
%functions of $m_{\tau}$, $m_b$, $m_c$, $\lambda_1$ and $\lambda_2$. 
%The hadronic parameters $\lambda_1$ and $\lambda_2$ are given by
%\beq
%K_b=-\frac{\lambda_1}{2m_b},\hspace{1.0cm}
%m_{B^*}^2-m_B^2=4\lambda_2,
%\eeq
%where $K_b$ is the kinetic energy of the $b$-quark inside $B$-meson
%and $m_B(m_{B^*})$ is the mass of $B(B^*)$-meson.
%Apart from $m_{\tau}=1.777$ GeV, the four parameters $m_b$, $m_c$,
%$\lambda_1$ and $\lambda_2$ are not independent. They are related by
%\beq
%m_B=m_b+\bar{\Lambda}-\frac{\lambda_1+3\lambda_2}{2m_b}+...,
%\hspace{1.0cm}
%m_D=m_c+\bar{\Lambda}-\frac{\lambda_1+3\lambda_2}{2m_c}+...,
%\eeq
%where $m_B=5.279$ GeV and $m_D=1.867$ GeV.
%We take the ranges of the three parameters $\bar{\Lambda}$, 
%$\lambda_1$ and $\lambda_2$ as follows \cite{GrLi}
%\beqar
%0.4<&\bar{\Lambda}&<0.6 ~~{\rm GeV}, \nonumber \\
%0<-&\lambda_1&<0.3 ~~{\rm GeV}^2, \nonumber \\
%0.11<&\lambda_2&<0.13 ~~{\rm GeV}^2. 
%\eeqar 
%
%We estimate $\Gamma_W$, $\Gamma_H$ and $\Gamma_I$ by varying the three
%parameters over the ranges given by the above equations.
%For $l=1,2$ and $q=c$($u$)
%\beqar
%\Gamma_W&=&0.5737 \pm 0.0699
%~(0.9726\pm0.0063)\nonumber \\
%\Gamma_H&=&0.6426 \pm 0.0873
%~(1.1009\pm0.0170)
%\eeqar
%and for $l=3$ and $q=c$($u$)
%\beqar
%\Gamma_W&=&0.1322\pm 0.0204
%~(0.3491\pm0.0187)\nonumber \\
%\Gamma_H&=&0.1639\pm 0.0294
%~(0.4301\pm0.0201)\nonumber \\
%\Gamma_I&=&0.0614\pm 0.0073
%~(0.1586\pm0.0162)
%\eeqar
%
%$m_b=4.7564\pm0.10345$
Since the species of the neutrinos cannot be distinguished by experiments
and the $\rp$ interactions allow the different kinds of the charged
lepton and the neutrino as decay products, we should sum the above decay
rates over neutrino species to compare with experimental data as follow
\beq
\Gamma_l^q\equiv\sum_{n=1}^{3}\Gamma_{ln}^q.
\eeq

%\section{experimental results}
%The semileptonic branching ratio of $B$-meson is defined as
%\beq
%B_{SL}=\frac{\Gamma(B\rightarrow X e \bar{\nu})}
%{\sum_{l}\Gamma(B\rightarrow X l \bar{\nu})+\Gamma_{had}+\Gamma_{rare}}
%\eeq
%where $\Gamma_{had}$ and $\Gamma_{rare}$ are the inclusive rates for
%hadronic and rare decays respectively. There are two kinds of measurements of
%the semileptonic branching ratio. One is performed at the $\Upsilon(4s)$
%resonance and the other at the $Z^0$ resonance. These two kinds of
%measurements are inconsistent \cite{Rich}. The average value at low energies is
%$B_{SL}=(10.23\pm0.39)\%$ and that at high energy is
%$B_{SL}=(11.23\pm0.34)\%$ using $B_{SL}=(\tau_B/\tau_b)B_{SL}(b)$ where
%$\tau_B=(1.60\pm0.03)$ps, $\tau_b=(1.56\pm0.03)$ps and
%$(b)$ indicates that this value refers to a mixture of $b$ hadrons.
%We take the average of these two measurements and inflate the error
%\cite{Neu}
%\beq
%B_{SL}=(10.80\pm0.51)\%.
%\eeq
%
%The semileptonic branching ratio for decays into $\tau$ lepton have been
%measured by the ALEPH and OPAL Collaborations. The weighted 
%average is 
%$B(B\rightarrow\tau\bar{\nu}X)=(2.68\pm0.28)\%$ \cite{Neu}. We obtain
%\beq
%\frac{B(B\rightarrow\tau\bar{\nu}X)}{B_{SL}}=(0.248\pm0.028)\%.
%\eeq

%The semileptonic branching ratios 
%due to $b\rightarrow u$ transition are not measured. 

%\section{product combinations constrained}
%\subsection{$\Gamma^u_{(1,2)}$ : $b ~ \rightarrow ~(e,\mu)~\bar{\nu}~X_u$}
For the process $b ~ \rightarrow ~e~\bar{\nu}~X_u$, 
the ratio of the decay rate in 
the MSSM without $R_p$ to that in the SM is given by
\beq
{\cal R}_1^u\equiv
\frac{\Gamma_1^u(\slash{R_p}~)}{\Gamma_1^u(SM)}=
%\frac{\left|V_{ub}\right|^2G_F^2m_b^5}{192\pi^3}
\left[\left|1+\ca_{11}^u\right|^2+\left|\ca_{12}^u\right|^2+
\left|\ca_{13}^u\right|^2\right]+
0.28~\left[\left|\cb_{11}^u\right|^2+\left|\cb_{12}^u\right|^2+
\left|\cb_{13}^u\right|^2\right].
\eeq
We will call the above ratio a $\slash{R_p}~$-ratio. 
This ratio is always bigger than 1 assuming
$\ca_{11}^u$ is real and positive.
Generally the above ratio could be smaller than 1.
% by $\slash{R_p}~$. 
We will consider this
possibility when we deal with the 
processes $ b ~ \rightarrow ~(e^-,\mu^-)~\bar{\nu}~X_c$ which are closely
related to the semileptonic branching ratio of $B$-meson.
For the process $b ~ \rightarrow ~\mu~\bar{\nu}~X_u$, 
the $\slash{R_p}~$-ratio is given by
\beq
{\cal R}_2^u\equiv
\frac{\Gamma_2^u(\slash{R_p}~)}{\Gamma_2^u(SM)}=
%\frac{\left|V_{ub}\right|^2G_F^2m_b^5}{192\pi^3}
\left[\left|\ca_{21}^u\right|^2+\left|1+\ca_{22}^u\right|^2+
\left|\ca_{23}^u\right|^2\right]+
0.28~\left[\left|\cb_{21}^u\right|^2+\left|\cb_{22}^u\right|^2+
\left|\cb_{23}^u\right|^2\right].
\eeq
The aboove two ratios could be largely affected by $\slash{R_p}~$ within the 
present bounds on $\llp{}{}$ and $\lplp{}{}$.

From the measurements of the ratio $\left|V_{ub}/V_{cb}\right|$ \cite{CLEO}
\beq
\left|V_{ub}/V_{cb}\right|=0.06~-~0.10,
\eeq
we can see that about 100 $\%$ of the SM rate is allowed in the process
$b\rightarrow u e^-(\mu^-) \bar{\nu}$ as a new physics contribution.
In Table \ref{haha1}, we list the combinations of couplings
whose present upper limits allow ${\cal R}_{(1,2)}^u$
to have values greater than 2,
assuming only one product of $R_p$-violating couplings is nonzero
and $\ca_{ll}^q$ is real and positive. 
For example, the bound $\lplp{132}{112}<4.8\times10^{-3}$ can allow
the ratio ${\cal R}_1^u$ to have the value of 9. 
From Table \ref{haha1}, we can see that
this large enhancement comes from the big differences between the values of
the CKM matrix elements.
This means that the experimental
determination of the ratio $\left|\frac{V_{ub}}{V_{cb}}\right|$ could be 
greatly affected by $\rp$.

There are no important contributions of $\cb_{(1,2)n}^u$ to 
${\cal R}_{(1,2)}^u$
taking into account the constraints coming from $B^0$-meson decays
into two charged leptons \cite{Lee}.
%The maximum values of ${\cal R}_1^u$ and ${\cal R}_2^u$ allowed by 
%these terms are 1.1 and 1.3 respectively. 
For example, let's assume that only $\llp{132}{113}$ which contributes
$\cb_{23}^u$ is not vanishing.
The upper bound on the product is 1.2$\times 10^{-3}$ without considering
the process $B^0 \rightarrow \mu^{\pm} \tau^{\mp}$. This upper bound gives
${\cal R}_2^u<2.2$. But the product is more strongly constrained by
the measurement of the process $B^0 \rightarrow \mu^{\pm} \tau^{\mp}$ :
$\llp{132}{113}<6.0\times 10^{-4}$. This gives ${\cal R}_2^u<1.3$.
The maximum values of ${\cal R}_1^u$ and ${\cal R}_2^u$ allowed by 
$\cb_{(1,2)n}^u$ are 1.1 and 1.3 respectively. 
%Therefore the contributions of the $\lambda$-type
%couplings to the semileptonic decays of $B$-meson would be negligible
%comparing with those of the $\lambda'$-type couplings within present bounds.

%\subsection{$\Gamma^u_{3}$ : $b ~ \rightarrow ~\tau~\bar{\nu}~X_u$}
For the process $b ~ \rightarrow ~\tau~\bar{\nu}~X_u$, 
the $\slash{R_p}~$-ratio 
is given by
\beqar
{\cal R}_3^u\equiv
\frac{\Gamma_3^u(\slash{R_p}~)}{\Gamma_3^u(SM)}&=&
%\frac{\left|V_{ub}\right|^2G_F^2m_b^5}{192\pi^3}&& 
\left[\left|\ca_{31}^u\right|^2+\left|\ca_{32}^u\right|^2+
\left|1+\ca_{33}^u\right|^2\right]+
%\nonumber \\
0.31~\left[\left|\cb_{31}^u\right|^2+\left|\cb_{32}^u\right|^2+
\left|R_3+\cb_{33}^u\right|^2\right] 
\nonumber \\
&-& 0.34~Re\left[\ca_{31}^{u*}\cb_{31}^{u}+\ca_{32}^{u*}\cb_{32}^u
+(1+\ca_{33}^{u*})(R_3+\cb_{33}^u) \right]
\left(\frac{m_{\tau}/1.8~{\rm GeV}}{m_b/4.8~{\rm GeV}}\right).
\eeqar
This process could be largely affected by $\rp$ (see Table \ref{haha1}).
There is no experimental evidence for this process at present.
%In this process we need some cautions in dealing with $\ca_{33}^u$ and 
%$\cb_{33}^u$ because there
%is a contribution from $R_3$. 
%Assuming only $\ca_{33}^u$ is not vanishing, the
%requirement ${\cal R}_3^u>2$ becomes
%\beq
%|1+\ca_{33}^u|^2+0.3080~|R_3|^2-0.3395~Re(R_3(1+\ca_{33}^{u*})) ~>~2.
%\eeq
%Neglecting $R_3$, we list the combinations of couplings whose maximum
%contributions satisfy ${\cal R}_3^u>2$, see Table \ref{haha1}.

%Assuming only $\cb_{33}^u$ is not vanishing, the
%requirement ${\cal R}_3^u>2$ becomes
%\beq
%1+0.3080~|R_3+\cb_{33}^u|^2-0.3395~Re(R_3+\cb_{33}^u) ~>~2.
%\eeq
%Assuming $\cb_{33}^u$ is real, we obtain $R_3+\cb_{33}^u>2.44$ from the above
%condition. Neglecting $R_3$,
%there is no combination whose maximum allowed value satisfy this.

%\subsection{$\Gamma^c_{(1,2)}$ : $b ~ \rightarrow ~(e,\mu)~\bar{\nu}~X_c$}
For the process $b ~ \rightarrow ~e~\bar{\nu}~X_c$, 
the $\slash{R_p}~$-ratio is given by
\beq
{\cal R}_1^c\equiv
\frac{\Gamma_1^c(\slash{R_p}~)}{\Gamma_1^c(SM)}=
%\frac{\left|V_{cb}\right|^2G_F^2m_b^5}{192\pi^3}
\left[\left|1+\ca_{11}^c\right|^2+\left|\ca_{12}^c\right|^2+
\left|\ca_{13}^c\right|^2\right]+
0.28~\left[\left|\cb_{11}^c\right|^2+\left|\cb_{12}^c\right|^2+
\left|\cb_{13}^c\right|^2\right].
\eeq
From the measurements of the semileptonic branching ratio ($B_{\rm SL}$)
\cite{Neu}
\beq
B_{\rm SL}=(10.80\pm0.51) \%,
\eeq
we can see that about 5 $\%$ of the SM rate is allowed in the process
$b~\rightarrow ~c~ e^-~ \bar{\nu}$ as a new physics contribution.
In Table \ref{haha2}, we list the combinations of couplings
whose present upper limits allow ${\cal R}_1^c$
to have values greater than 1.1
assuming only one product of $\slash{R_p}~$-violating couplings is nonzero
and $\ca_{ll}^q$ is real and positive. 
%The contributions of $\cb_{1n}^c$ are negligible.

For the process $b ~ \rightarrow ~\mu~\bar{\nu}~X_c$ the 
$\slash{R_p}~$-ratio is given by
\beq
{\cal R}_2^c\equiv
\frac{\Gamma_2^c(\slash{R_p}~)}{\Gamma_2^c(SM)}=
%\frac{\left|V_{cb}\right|^2G_F^2m_b^5}{192\pi^3}
\left[\left|\ca_{21}^c\right|^2+\left|1+\ca_{22}^c\right|^2+
\left|\ca_{23}^c\right|^2\right]+
0.28~\left[\left|\cb_{21}^c\right|^2+\left|\cb_{22}^c\right|^2+
\left|\cb_{23}^c\right|^2\right].
\eeq
We list the combinations of couplings similar to the case of
$b ~ \rightarrow ~e~\bar{\nu}~X_c$ in Table \ref{haha2}.
We also observe that 
the contributions of $\cb_{(1,2)n}^c$ are negligible taking 
into account the constraints coming from $B^0$-meson decays
into two charged leptons.

From the considerations of the processes
$b\rightarrow e (\mu) \bar{\nu} X_{u,c}$, we can see that
the contributions of the $\lambda$-type
couplings to the semileptonic decays of $B$-meson would be negligible
comparing with those of the $\lambda'$-type couplings within present bounds.

%\subsection{Semileptonic branching ratio}
If we loose the assumption that $\ca_{11}^c$ and $\ca_{22}^c$ are real and
positive, $\slash{R_p}~$ can decrease the
semileptonic branching ratio. For example, let's
assume that only $\lplp{132}{122}$ 
which contributes to $\ca_{11}^c$ is
nonzero. Since the upper bound on the {\it magnitude} on this combination is 
$4.8\times 10^{-3}$, it can decrease ${\cal R}_1^c$ by 0.3. 
In fact we obtain
\beqar
 0.7 <& {\cal R}_1^c& < 1.5, \nonumber \\
 0.7 <& {\cal R}_2^c& < 1.7. 
\eeqar
This implies that it is possible to explain 
the gap between the measured and the expected values for the
semileptonic branching ratio of $B$-meson by $\rp$.

$\rp$ could results in the lepton non-universality.
The semileptonic branching ratios $b\rightarrow e \nu X$ and
$b\rightarrow \mu \nu X$ measured by the L3 Collaborations \cite{L3} are
\beqar
B(b\rightarrow e \nu X)&=&(10.89\pm0.55)~\%, \nonumber \\
B(b\rightarrow \mu \nu X)&=&(10.82\pm0.61)~\%.
% \nonumber \\
%\frac{B(b\rightarrow e \nu X)}{B(b\rightarrow \mu \nu X)}
%&=&(1.00\pm0.08), \nonumber \\
%\frac{B(b\rightarrow \mu \nu X)}{B(b\rightarrow e \nu X)}
%&=&(0.99\pm0.08). 
\eeqar
Considering this lepton universality measurements under the assumption that
$\rp$ does not contribute to these two semileptonic 
branching ratios simultaneously, we can derive $1\sigma$ bounds on single and
some products of $\rp$
couplings slightly stronger than previous ones, see Table \ref{haha3}. 

%\subsection{$\Gamma^c_{3}$ : $b ~ \rightarrow ~ \tau~\bar{\nu}~X_c$}
For the process $b ~ \rightarrow ~ \tau~\bar{\nu}~X_c$, 
the $\rp$-ratio is given by
\beqar
{\cal R}_3^c\equiv
\frac{\Gamma_3^c(\slash{R_p}~)}{\Gamma_3^c(SM)}&=&
%\frac{\left|V_{cb}\right|^2G_F^2m_b^5}{192\pi^3}&& 
\left[\left|\ca_{31}^c\right|^2+\left|\ca_{32}^c\right|^2+
\left|1+\ca_{33}^c\right|^2\right]+
%\nonumber \\
0.31~\left[\left|\cb_{31}^c\right|^2+\left|\cb_{32}^c\right|^2+
\left|R_3+\cb_{33}^c\right|^2\right] 
\nonumber \\
&-& 0.35~Re\left[\ca_{31}^{c*}\cb_{31}^{c}+\ca_{32}^{c*}\cb_{32}^c
+(1+\ca_{33}^{c*})(R_3+\cb_{33}^c) \right]
\left(\frac{m_{\tau}/1.8~{\rm GeV}}{m_b/4.8~{\rm GeV}}\right).
\eeqar
Using the SM prediction for the branching ratio 
$B(\bar{B}\rightarrow \tau \bar{\nu} X)=2.30\pm0.25 \%$ \cite{Falk} and
the experimental results of the branching ratio, $2.68\pm0.28\%$ \cite{Neu}
and assuming
$\frac{B({b \rightarrow \tau \bar{\nu}~X (\slash{R_p}~)})}
{B({b \rightarrow \tau \bar{\nu}~X (SM)})} \approx {\cal R}_3^c$, we obtain
\beq
{\cal R}_3^c < 1.34~~(1\sigma).
\eeq
We find that all the constraints on the products of $\rp$ couplings
coming from the measurement of the branching ratio of the process
$\bar{B}\rightarrow \tau \bar{\nu} X$
are weaker than previous ones neglecting $R_3$. We list some competitive upper
bounds on $\rp$ couplings from
$\bar{B}\rightarrow \tau \bar{\nu} X$ in Table \ref{haha4}.
One of them ($\lambda'_{333}$) was firstly given in Ref. \cite{Feng}.

To see the effects of $\rp$ on the determination of the upper bound
on $r=\tan\beta/m_{H^{\pm}}$, we assume only $\ca_{33}^c$ is not vanishing.
The condition ${\cal R}_3^c<1.34$ becomes
\beq
|1+\ca_{33}^c|^2+0.31~|R_3|^2-0.35~Re(R_3(1+\ca_{33}^{c*})) ~<~1.34.
\eeq
Assuming $\ca_{33}^c$ is real and 
using the present bound $|\ca_{33}^c|<1.0\times10^{-1}$ and 
we find
\beqar
r&<&0.54~~{\rm GeV}^{-1}~~~~{\rm for}~~~~~ \ca_{33}^c=-0.1, \nonumber \\
r&<&0.52~~{\rm GeV}^{-1}~~~~{\rm for}~~~~~ \ca_{33}^c=0, \nonumber \\
r&<&0.51~~{\rm GeV}^{-1}~~~~{\rm for}~~~~~ \ca_{33}^c=0.1. 
\eeqar
% and we varies $m_b^Y$ as $0.9~m_b<m_b^Y<1.1~m_b$. 
We obtain slight weak upper bounds on $r$
than that of Ref. \cite{GrLi} : $r<0.51~{\rm GeV}^{-1}$. We observe
the effects of $\slash{R_p}~$ on the bound on $r$ are negligible.

%Assuming only $\cb_{33}^c$ is not vanishing, the
%condition ${\cal R}_3^c<1.341$ becomes
%\beq
%1+0.3099~|R_3+\cb_{33}^c|^2-0.3470~Re(R_3+\cb_{33}^c) ~<~1.341.
%\eeq
%Assuming $\cb_{33}^c$ is real we obtain
%\beq
%-0.76 <R_3+\cb_{33}^c < 2.39.
%\eeq

In conclusion, 
we investigate the effects of $\slash{R_p}~$ on the semileptonic decays of
$B$-meson in the MSSM with explicit $R$-parity violation.
We find that $\rp$ has large effects on the experimental determination of the
ratio $\left|\frac{V_{ub}}{V_{cb}}\right|$ and could decrease the semileptonic
branching ratio. The effects of $\lambda$-type couplings are negligible
compared with those of $\lambda'$-type couplings. Also are derived 
the bounds on single and some products of $\rp$ couplings slightly stronger
than previous ones.
We observe that the effects of $\rp$ on the bound on $\tan\beta/m_{H^{\pm}}$ 
are negligible.

\section*{acknowledgements}
We thank K. Choi and E. J. Chun for their helpful remarks.
This work was supported in part by KAIST Basic Science Research Program
(J.S.L.).

\begin{table}
\caption{\label{haha1}
Maximally allowed $\slash{R_p}~$-ratios and the list of combinations
whose present upper bounds are weak enough to allow the ratios to have the
value greater than 2 for the processes $b~\rightarrow~e_l~\bar{\nu}~X_u$. 
We assume that
only one combination is nonzero and $\ca_{ll}^u$ is real and positive.
We use the magnitudes of the CKM matrix elements and the masses of
the squarks and the sleptons as shown in Eq. (6).
%(a):$t$-decay \cite{agagra}, (b):$K^+$-decay \cite{agagra}, (c):$Z$ decay
%width \cite{Zdecay}
}
\begin{tabular}{lclc}
%%%%%%%%%%%%%%%%%%%%%%%%%%%%%%%%%%%%%%%%%%%%%%%%%%%%%%%%%%%%%%%%%%%%%%%%
Processes & Combinations & Upper Bound $\times$ CKM & 
Maximum of $\slash{R_p}~$-ratio \\
%%%%%%%%%%%%%%%%%%%%%%%%%%%%%%%%%%%%%%%%%%%%%%%%%%%%%%%%%%%%%%%%%%%%%%%%
\hline
$b~\rightarrow~e~\bar{\nu}~X_u$ 
& $\lplp{132}{112}$ & $4.8\times10^{-3}\times(V_{ud}/V_{ub})^{\rm a,b}$ & 9.0 \\
& $\lplp{132}{122}$ & $4.8\times10^{-3}\times(V_{us}/V_{ub})^{\rm a,b}$ & 2.1 \\
& $\lplp{232}{112}$ & $5.3\times10^{-3}\times(V_{ud}/V_{ub})^{\rm c,b}$ & 6.0 \\
& $\lplp{233}{113}$ & $5.3\times10^{-3}\times(V_{ud}/V_{ub})^{\rm c,b}$ & 6.0 \\
& $\lplp{332}{112}$ & $3.1\times10^{-3}\times(V_{ud}/V_{ub})^{\rm c,b}$ & 2.7 \\
& $\lplp{333}{113}$ & $3.1\times10^{-3}\times(V_{ud}/V_{ub})^{\rm c,b}$ & 2.7 \\
%%%%%%%%%%%%%%%%%%%%%%%%%%%%%%%%%%%%%%%%%%%%%%%%%%%%%%%%%%%%%%%%%%%%%%%%
\hline
$b~\rightarrow~\mu~\bar{\nu}~X_u$
& $\lplp{131}{211}$ & $3.1\times10^{-3}\times(V_{ud}/V_{ub})^{\rm d,b}$ & 2.7 \\
& $\lplp{132}{212}$ & $4.8\times10^{-3}\times(V_{ud}/V_{ub})^{\rm a,b}$ & 5.0 \\
& $\lplp{231}{211}$ & $2.6\times10^{-3}\times(V_{ud}/V_{ub})^{\rm e,b}$ & 4.5 \\
& $\lplp{232}{212}$ & $5.3\times10^{-3}\times(V_{ud}/V_{ub})^{\rm c,b}$ & 10.4 \\
& $\lplp{232}{222}$ & $5.3\times10^{-3}\times(V_{us}/V_{ub})^{\rm c,b}$ & 2.3 \\
& $\lplp{233}{213}$ & $5.3\times10^{-3}\times(V_{ud}/V_{ub})^{\rm c,b}$ & 10.4 \\
& $\lplp{233}{223}$ & $5.3\times10^{-3}\times(V_{us}/V_{ub})^{\rm c,b}$ & 2.3 \\
& $\lplp{331}{211}$ & $3.1\times10^{-3}\times(V_{ud}/V_{ub})^{\rm c,b}$ & 2.7 \\
& $\lplp{332}{212}$ & $3.1\times10^{-3}\times(V_{ud}/V_{ub})^{\rm c,b}$ & 2.7 \\
& $\lplp{333}{213}$ & $3.1\times10^{-3}\times(V_{ud}/V_{ub})^{\rm c,b}$ & 2.7 \\
%%%%%%%%%%%%%%%%%%%%%%%%%%%%%%%%%%%%%%%%%%%%%%%%%%%%%%%%%%%%%%%%%%%%%%%%
\hline
$b~\rightarrow~\tau~\bar{\nu}~X_u$
& $\lplp{131}{311}$ & $3.1\times10^{-3}\times(V_{ud}/V_{ub})^{\rm d,b}$ & 2.7 \\
& $\lplp{132}{312}$ & $4.8\times10^{-3}\times(V_{ud}/V_{ub})^{\rm a,b}$ & 5.0 \\
& $\lplp{231}{311}$ & $2.6\times10^{-3}\times(V_{ud}/V_{ub})^{\rm e,b}$ & 2.2 \\
& $\lplp{232}{312}$ & $5.3\times10^{-3}\times(V_{ud}/V_{ub})^{\rm c,b}$ & 5.8 \\
& $\lplp{233}{313}$ & $5.3\times10^{-3}\times(V_{ud}/V_{ub})^{\rm c,b}$ & 5.8 \\
& $\lplp{331}{311}$ & $3.1\times10^{-3}\times(V_{ud}/V_{ub})^{\rm c,b}$ & 5.3 \\
& $\lplp{332}{312}$ & $3.1\times10^{-3}\times(V_{ud}/V_{ub})^{\rm c,b}$ & 5.3 \\
& $\lplp{333}{313}$ & $3.1\times10^{-3}\times(V_{ud}/V_{ub})^{\rm c,b}$ & 5.3 \\
%%%%%%%%%%%%%%%%%%%%%%%%%%%%%%%%%%%%%%%%%%%%%%%%%%%%%%%%%%%%%%%%%%%%%%%%
\end{tabular}
(a): $t$-decay ($2\sigma$)\cite{agagra}, 
(b): $K^+$-decay (90 \% C.L.) \cite{agagra,choroy}, 
(c): $Z$ decay width ($1\sigma$) \cite{Zdecay}, 
(d): Atomic parity violation and $eD$ asymmetry  ($1\sigma$) \cite{han},
(e): $\nu_{\mu}$ deep-inelastic scattering  ($2\sigma$) \cite{han}.
\end{table}

\begin{table}
\caption{\label{haha2}
Maximally allowed $\slash{R_p}~$-ratios and the list of combinations
whose present upper bounds are weak enough to allow the ratios to have the
value greater than 1.1 for the processes $b~\rightarrow~e_l~\bar{\nu}~X_c$. 
We assume that
only one combination is nonzero and $\ca_{ll}^c$ is real and positive.
We use the magnitudes of the CKM matrix elements and the masses of
the squarks and the sleptons as shown in Eq. (6).
}
\begin{tabular}{lclc}
%%%%%%%%%%%%%%%%%%%%%%%%%%%%%%%%%%%%%%%%%%%%%%%%%%%%%%%%%%%%%%%%%%%%%%%%
Processes & Combinations & Upper Bound $\times$ CKM & 
Maximum of $\slash{R_p}~$-ratio \\
%%%%%%%%%%%%%%%%%%%%%%%%%%%%%%%%%%%%%%%%%%%%%%%%%%%%%%%%%%%%%%%%%%%%%%%%
\hline
$b~\rightarrow~e~\bar{\nu}~X_c$ 
& $\lplp{131}{121}$    & $3.1\times10^{-3}\times(V_{cs}/V_{cb})^{\rm a,b}$ & 1.2 \\
& $|\lambda'_{131}|^2$ & $6.8\times10^{-2}\times(V_{cb}/V_{cb})^{\rm a}$ & 1.2 \\
& $\lplp{132}{122}$    & $4.8\times10^{-3}\times(V_{cs}/V_{cb})^{\rm c,b}$ & 1.4 \\
& $|\lambda'_{132}|^2$ & $1.6\times10^{-1}\times(V_{cb}/V_{cb})^{\rm c}$ & 1.5 \\
%%%%%%%%%%%%%%%%%%%%%%%%%%%%%%%%%%%%%%%%%%%%%%%%%%%%%%%%%%%%%%%%%%%%%%%%
\hline
$b~\rightarrow~\mu~\bar{\nu}~X_c$
& $\lplp{231}{221}$    & $2.6\times10^{-3}\times(V_{cs}/V_{cb})^{\rm d,b}$ & 1.2 \\
& $|\lambda'_{231}|^2$ & $4.8\times10^{-2}\times(V_{cb}/V_{cb})^{\rm d}$ & 1.2 \\
& $\lplp{232}{222}$    & $5.3\times10^{-3}\times(V_{cs}/V_{cb})^{\rm e,b}$ & 1.4 \\
& $|\lambda'_{232}|^2$ & $1.9\times10^{-1}\times(V_{cb}/V_{cb})^{\rm e}$ & 1.7 \\
& $\lplp{233}{223}$    & $5.3\times10^{-3}\times(V_{cs}/V_{cb})^{\rm e,b}$ & 1.4 \\
& $|\lambda'_{233}|^2$ & $1.9\times10^{-1}\times(V_{cb}/V_{cb})^{\rm e}$ & 1.7 \\
%%%%%%%%%%%%%%%%%%%%%%%%%%%%%%%%%%%%%%%%%%%%%%%%%%%%%%%%%%%%%%%%%%%%%%%%
\end{tabular}
(a): Atomic parity violation and $eD$ asymmetry  ($1\sigma$) \cite{han},
(b): $K^+$-decay (90 \% C.L.) \cite{agagra,choroy}, 
(c): $t$-decay ($2\sigma$)\cite{agagra}, 
(d): $\nu_{\mu}$ deep-inelastic scattering  ($2\sigma$) \cite{han},
(e): $Z$ decay width ($1\sigma$) \cite{Zdecay}. 
\end{table}

\begin{table}
\caption{\label{haha3}
Upper bounds on the magnitudes of $\rp$ couplings from the lepton universality
in the semileptonic decays of $B$-meson.
We use the magnitudes of the CKM matrix elements and the masses of
the squarks and the sleptons as shown in Eq. (6).
}
\begin{tabular}{lcll}
%%%%%%%%%%%%%%%%%%%%%%%%%%%%%%%%%%%%%%%%%%%%%%%%%%%%%%%%%%%%%%%%%%%%%%%%
Processes & Combinations Constrained & Upper Bound& 
Previous Bound \\
%%%%%%%%%%%%%%%%%%%%%%%%%%%%%%%%%%%%%%%%%%%%%%%%%%%%%%%%%%%%%%%%%%%%%%%%
\hline
$b~\rightarrow~e~\bar{\nu}~X_c$ 
& $\lplp{131}{121}$ & $1.1\times10^{-3}$& $3.1\times10^{-3~~\rm a,b}$\\
& $\lambda'_{131}$  & $1.6\times10^{-1}$& $2.6\times10^{-1~~\rm a}$\\
& $\lplp{132}{122}$ & $1.1\times10^{-3}$& $4.8\times10^{-3~~\rm c,b}$\\
& $\lambda'_{132}$  & $1.6\times10^{-1}$& $4.0\times10^{-1~~\rm c}$\\
%%%%%%%%%%%%%%%%%%%%%%%%%%%%%%%%%%%%%%%%%%%%%%%%%%%%%%%%%%%%%%%%%%%%%%%%
\hline
$b~\rightarrow~\mu~\bar{\nu}~X_c$
& $\lplp{231}{221}$& $1.1\times10^{-3}$& $2.6\times10^{-3~~\rm d,b}$ \\
& $\lambda'_{231}$ & $1.6\times10^{-1}$& $2.2\times10^{-1~~\rm d}$\\
& $\lplp{232}{222}$& $1.1\times10^{-3}$& $5.3\times10^{-3~~\rm e,b}$\\
& $\lambda'_{232}$ & $1.6\times10^{-1}$& $4.4\times10^{-1~~\rm e}$\\
& $\lplp{233}{223}$& $1.1\times10^{-3}$& $5.3\times10^{-3~~\rm e,b}$\\
& $\lambda'_{233}$ & $1.6\times10^{-1}$& $4.4\times10^{-1~~\rm e}$\\
%%%%%%%%%%%%%%%%%%%%%%%%%%%%%%%%%%%%%%%%%%%%%%%%%%%%%%%%%%%%%%%%%%%%%%%%
\end{tabular}
(a): Atomic parity violation and $eD$ asymmetry  ($1\sigma$) \cite{han},
(b): $K^+$-decay (90 \% C.L.) \cite{agagra,choroy}, 
(c): $t$-decay ($2\sigma$)\cite{agagra}, 
(d): $\nu_{\mu}$ deep-inelastic scattering  ($2\sigma$) \cite{han},
(e): $Z$ decay width ($1\sigma$) \cite{Zdecay}. 
\end{table}

\begin{table}
\caption{\label{haha4}
Upper bounds on the magnitudes of $\rp$ couplings from 
$\bar{B}\rightarrow \tau \bar{\nu} X$. 
We use the magnitudes of the CKM matrix elements and the masses of
the squarks and the sleptons as shown in Eq. (6).
}
\begin{tabular}{lcll}
%%%%%%%%%%%%%%%%%%%%%%%%%%%%%%%%%%%%%%%%%%%%%%%%%%%%%%%%%%%%%%%%%%%%%%%%
Processes & Combinations Constrained & Upper Bound& 
Previous Bound \\
%%%%%%%%%%%%%%%%%%%%%%%%%%%%%%%%%%%%%%%%%%%%%%%%%%%%%%%%%%%%%%%%%%%%%%%%
\hline
$\bar{B}\rightarrow \tau \bar{\nu} X$ 
& $\lplp{331}{321}$ & $4.4\times10^{-3}$& $3.1\times10^{-3~~\rm a,b}$\\
& $\lambda'_{331}$  & $3.2\times10^{-1}$& $2.6\times10^{-1~~\rm a}$\\
& $\lplp{332}{322}$ & $4.4\times10^{-3}$& $3.1\times10^{-3~~\rm a,b}$\\
& $\lambda'_{332}$  & $3.2\times10^{-1}$& $2.6\times10^{-1~~\rm a}$\\
& $\lplp{333}{323}$ & $4.4\times10^{-3}$& $3.1\times10^{-3~~\rm a,b}$\\
& $\lambda'_{333}$  & $3.2\times10^{-1~~\rm c}$& $2.6\times10^{-1~~\rm a}$\\
%%%%%%%%%%%%%%%%%%%%%%%%%%%%%%%%%%%%%%%%%%%%%%%%%%%%%%%%%%%%%%%%%%%%%%%%
\end{tabular}
(a): $Z$ decay width ($1\sigma$) \cite{Zdecay}, 
(b): $K^+$-decay (90 \% C.L.) \cite{agagra,choroy},
(c): $\bar{B}\rightarrow \tau \bar{\nu} X$ \cite{Feng}.
\end{table}


\begin{references}
\bibitem{ago}
\plb{J. Ellis et al.}{150}{142}{1985};
\plb{G. G. Ross and J. W. F. Valle}{151}{375}{1985};
\npb{S. Dawson}{261}{297}{1985};
\plb{S. Dimopoulos and L. Hall}{207}{210}{1987}.
\bibitem{Bfac}
T. Nakada, PSI-PR-96-22, hep-ex/9609015. F. Grancagnolo, INFN-AE-90-07.
CLEO Collaborations, CLNS-94-1277. J. N. Butler, FERMILAB-PUB-95-363, Nucl.
Instrum. Meth. {\bf A368}, 145 (1995), D. E. Kaplan, UW/PT 97-5, hep-ph/9703347.
\bibitem{han}
\prd{V. Barger, G. F. Giudice and T. Han}{40}{2987}{1989}.
\bibitem{beta}
\prd{R. N. Mohapatra}{34}{3457}{1986};
\prl{M. Hirsch, H. V. Klapdor--Kleingrothaus and S. G.
Kovalenko}{75}{17}{1995}.
\bibitem{numass}
\npb{R. M. Godbole, P. Roy and X. Tata}{401}{67}{1993}.
\bibitem{agagra}
\prd{K. Agashe, M. Graesser}{54}{4445}{1996}.
\bibitem{Zdecay}
G. Bhattacharyya, J. Ellis, and K. Sridhar, Mod. Phys. $\bf{A10}$, 1583
(1995), \plb{G. Bhattacharyya, D. Choudhury and K. Sridhar}{355}{193}{1995}.
\bibitem{bha}
G. Bhattacharyya, preprint IFUP-TH-52/96, hep-ph/9608415.
\bibitem{chahu}
\plb{M. Chaichian and K. Huitu}{384}{157}{1996}.
\bibitem{choroy}
\plb{D. Choudhury and P. Roy}{378}{153}{1996}.
\bibitem{Lee}
J. Jang, J. K. Kim and J. S. Lee, KAIST-TH 97/02, hep-ph/9701283, to appear in
Phys. Rev. D.
\bibitem{Feng}
\prl{J. Erler, J. L. Feng and N. Polonsky}{78}{3063}{1997}, 
\bibitem{Ko}
J. E. Kim, P. Ko and D. Lee, SNUTP 96-073, hep-ph/9701381.
\bibitem{Choi}
\prd{K. Choi, E. J. Chun and J. S. Lee}{55}{R3924}{1997}, 
\bibitem{hasu}
\npb{L. J. Hall and M. Suzuki}{231}{419}{1984};
F. Vissani, hep-ph/9607423.
\bibitem{GrLi}
\plb{Y. Grossman and Z. Ligeti}{332}{373}{1994},
\plb{J. Kalinowski}{245}{201}{1990}.
\bibitem{PDG}
\prd{Review of Particle Properties}{54}{1}{1996}.
\bibitem{Falk}
\plb{A. F. Falk, Z. Ligeti, M. Neubert and Y. Nir}{326}{145}{1994}.
%\bibitem{Rich}
%J. D. Richman, hep-ex/9701014
\bibitem{CLEO}
CLEO Collab., J. Bartlet {\it et al}, Phys. Rev. Lett. $\bf{71}$, 4111 (1993).
\bibitem{Neu}
M. Neubert, CERN-TH/97-19, hep-ph/9702310
%\bibitem{dross}
%\npb{H. Dreiner and G. G. Ross}{410}{188}{1993}.
\bibitem{L3}
L3 Collab., M. Acciarri {\it et al}, Z. Phys. $\bf{C 71}$, 379 (1996).
%\bibitem{Zdecay}
%G. Bhattacharyya, J. Ellis, and K. Sridhar, Mod. Phys. $\bf{A10}$, 1583 (1995).
\end{references}
\end{document}